\documentclass{article}
\usepackage[utf8]{inputenc} 

\usepackage{geometry} 
\usepackage{graphicx} 
\usepackage{amsmath}
\usepackage{amssymb}
\usepackage{booktabs} 
\usepackage{array} 
\usepackage{paralist} 
\usepackage{verbatim} 
\usepackage{subfig} 
\usepackage{float}
\usepackage{fancyhdr} 
\usepackage{sectsty}
\allsectionsfont{\sffamily\mdseries\upshape} 
\usepackage[nottoc,notlof,notlot]{tocbibind} 
\usepackage[titles,subfigure]{tocloft} 

\newtheorem{example}{Example}
\newtheorem{remark}{Remark}

\newcommand{\bA}{{\mathbb{A}}}
\newcommand{\bB}{{\mathbb{B}}}
\newcommand{\bL}{{\mathbb{L}}}
\newcommand{\cA}{{\mathcal{A}}}

\newcommand{\sgn}{\text{sgn}}

\title{Non-markovian mesoscopic dissipative dynamics\\ of open quantum spin chains}

\author{F. Benatti$^{1,2}$, F. Carollo$^{1,2}$, R. Floreanini$^2$, H. Narnhofer$^{3}$\\
\small ${}^1$Dipartimento di Fisica, Universit\`a di Trieste, Strada Costiera 11, I-34151, Trieste, Italy\\
\small ${}^2$Istituto Nazionale di Fisica Nucleare, Sezione di Trieste, Strada Costiera 11, I-34151, Trieste, Italy\\
\small ${}^3$Institut f\"ur Theoretische Physik, Boltzmanngasse 5, A-1090, Vienna, Austria}

\date{\null}

\begin{document}
\maketitle

\begin{abstract}
We study the dissipative dynamics of $N$ quantum spins with Lindblad
generator consisting of operators scaling as fluctuations, namely with
the inverse square-root of $N$.
In the large $N$ limit, the microscopic dissipative time-evolution
converges to a non-Markovian unitary dynamics on strictly local
operators, while at the mesoscopic level of fluctuations it gives rise
to a dissipative non-Markovian dynamics.
The mesoscopic time-evolution is Gaussian and exhibits either a stable
or an unstable asymptotic character; furthermore, the mesoscopic
dynamics builds correlations among fluctuations that survive in time
even when the original microscopic dynamics is unable to correlate
local observables.
\end{abstract}

\section{Introduction}
In many physical situations concerning many-body quantum systems with $N$ microscopic components, the relevant observables are not those referring to single constituents, rather the collective ones consisting of suitably scaled sums of microscopic operators. Among them, one usually considers  macroscopic averages that scale as the inverse of $N$: these loose all quantum properties in the large $N$ limit thereby providing a classical description of the collective features of many body quantum systems.

Another class of relevant collective observables are the so-called quantum fluctuations: they account for the variations of microscopic quantities around their averages computed with respect to a chosen reference state. 
In analogy to classical fluctuations, they scale with the inverse square root of $N$ so that, unlike macroscopic observables, they can retain quantum features in the large $N$ limit \cite{BOOK,GVVNCCL,GVVDOF}. Indeed, whenever the reference microscopic state presents no long-range correlations in the large $N$ limit, the fluctuations behave as bosonic operators; furthermore, from the microscopic state there emerges a Gaussian state over the corresponding bosonic Canonical Commutation Relation (CCR) algebra.
These collective observables describe a mesoscopic physical scale on which many-body behaviours are in between the purely quantum behaviour of microscopic observables and the purely classical one of  macroscopic observables \cite{HT}. 

Due to the large number of degrees of freedom, the analytic description of the time-evolution of many-body systems is usually impossible. Nevertheless, in many situations \cite{BOOK,Thirr}, mean-field approximations lead to dynamics amenable to quantitative and qualitative considerations.
Mean-field unitary dynamics have already been extensively studied at the level of fluctuations operators \cite{BOOK,HT,NARN}. In the following  we consider the case of a mean-field dissipative dynamics of an open quantum spin chain weakly coupled to an environment through collective operators scaling with the inverse square-root of the number of individual constituents, as in the Dicke model of matter-light interaction \cite{HL,sew,sew2}.

We study the large $N$ limit of the dissipative time-evolution at the microscopic scale of local observables, that is for fixed number of sites, and at the mesoscopic level of quantum fluctuations. 
These two scenarios look quite different: when $N\to\infty$, local observables evolve according to a Hamiltonian despite the dynamics being purely dissipative for each finite $N$. This Hamiltonian is time-dependent if mean-field operators tend to macroscopic mean magnetisations that vary with time. The same fate occurs to quantum fluctuations; however, their mesoscopic dynamics is a dissipative family of completely positive maps that send Gaussian states into Gaussian states. 
In the time-dependent case, the generators of both local and mesoscopic time-evolutions depend explicitly not only on the running time $t$, but also on the initial time $t_0$; therefore, when $N\to\infty$, the microscopic dynamics, which composes as a one-parameter semigroup, gives rise to a non-Markovian time-evolution on the microscopic as well as on the mesocopic level.

It turns out that the dynamics of the mean magnetisations possesses a stable and a unstable asymptotic point; correspondingly, the semigroup of mesoscopic, dissipative Gaussian maps has an attracting asymptotic Gaussian state or none at all.
In the former case, the asymptotic Gaussian state embodies correlations among 
fluctuations that are built by the mesoscopic collective dynamics even when it emerges from a microscopic time-evolution that is unable to correlate local observables.

\section{The Model}

We consider a quantum spin chain consisting of a doubly infinite lattice, with a two-level system at each of its sites described by the algebra $\cA_2$ generated by
$s_\mu=\sigma_\mu/2$, $\mu=1,2,3$, and $s_0=\frac{{\bf 1}}{2}$, with $\sigma_\mu$ the Pauli matrices so that 
\begin{equation}
\label{Pauli}
[s_\mu\,,\,s_\nu]=\,i\,\epsilon_{\mu\nu\delta}\,s_\delta\ .
\end{equation} 
The different sites will be labeled by integer numbers $k\in\mathbb{Z}$ so that $s_\mu^{(k)}$ will denote the spin operator $s_\mu$ pertaining to the $k$-th site. As customary for quantum spin chains, the algebra $\cA$ describing the infinite chain is chosen to be the so-called \textit{quasi-local} $C^*$ algebra that arises from the norm closure of the algebra generated by spin operators differing from the identity only at finitely, but arbitrarily  many sites \cite{BraRob}.

On this algebra $\cA$, a state $\omega$ is a positive, linear, normalized functional $\omega:\cA\to\mathbb{C}$ assigning to each $a\in\cA$ its expectation value $\omega(a)$. 
Notice that the mean-values of the operators $s_\mu$ satisfy
\begin{equation}
\label{Pauli2}
|\omega(s_\mu)|\leq\frac{1}{2}\ .
\end{equation}

We assume the state to be translation invariant, namely $\displaystyle
\omega\left(\tau(a)\right)=\omega(a)$ for all $a\in\cA$,
where $\tau$ is the translation automorphism moving spin operators from one site to the one at its right, $\tau(s_\mu^{(k)})=s_\mu^{(k+1)}$. Furthermore, we shall ask the state to  be clustering, namely to obey
\begin{equation}
\lim_{|k|\to\infty}\omega\left(\tau^{k}\left(a\right)b\right)=\omega(a)\omega(b)\qquad
\forall a\,,b\in\cA\ .
\end{equation}
Physically, this means that $\omega$ carries no correlations between spin observables supported by lattice regions far away from each other.

The relevant physical properties of many-body systems are encoded not in microscopic, local observables, rather in coarse-grained observables involving all their constituents and 
giving access to the statistical properties of the system as a whole. 
Most natural among these collective observables are the averages over microscopic observables like, for instance, the mean magnetisation along the $\mu$-axis,
\begin{equation}
\label{avN}
S_\mu^{(N)}=\frac{1}{N}\sum_{k=1}^Ns_\mu^{(k)}\ .
\end{equation}
One then studies the convergence of $S_\mu^{(N)}$ as $N\to\infty$: it turns out that, with respect to translation invariant clustering states, the so-called weak limit of these averages provides multiples of the identity operator:
\begin{equation}
\label{av}
S_\mu=w-\lim_{N\to\infty}S_\mu^{(N)}=\omega\left(s_\mu\right){\bf 1}\ ,
\end{equation}
the weak convergence of a sequence $\{c_N\}_{N\in\mathbb{N}}$ of operators $c_N\in\mathcal{A}$ to an operator $c$ meaning that
\begin{equation}
\label{wlim}
w-\lim_{N\to\infty}c_N=c\Leftrightarrow\lim_{N\to\infty}\omega\left(a^\dag c_N b\right)=\omega(a^\dag\,c\, b)\qquad \forall\ a\,,b\in\mathcal{A}\ .
\end{equation}
The averages $S_\mu$ commute among themselves and are thus appropriately associated with the "macroscopic" classical scale of the quantum spin chain defined by the 
the state $\omega$ \footnote{In technical terms, the weak-limits correspond to limits within the so-called GNS-representation defined by the state $\omega$ \cite{BraRob}\label{foot1}\ .}.

However, while a scaling with inverse powers of $N$ is necessary in order to arrive at a sound collective description, $1/N$ is not the only one possible. Other collective operators can be defined by the weaker scaling $1/\sqrt{N}$: this is typical of  fluctuations of classical stochastic variables around their mean-values with respect to a given probability distribution.
Analogously, for a quantum spin chain, the fluctuation of a single spin operator is defined 
by:
\begin{equation}
F(s_\mu)=m-\lim_{N\to\infty}F_N(s_\mu)\ ,\quad F_N(s_\mu)=\frac{1}{\sqrt{N}}\sum_{k=1}^{N}\left(s_\mu^{(k)}-\omega(s_\mu)\right)\ ,
\label{fluc}
\end{equation} 
where the $m-lim$ denotes the mesoscopic limit and has to be understood in the \emph{quantum central limit} sense \cite{GVVNCCL}.
Before explaining what this concretely means, let us first observe that such a limit is expected to preserve quantum features. Indeed, since spins at different sites commute, from \eqref{Pauli} it follows that
$$
\left[F_N(s_\mu),F_N(s_\nu)\right]=i\epsilon_{\mu\nu\gamma}\,S_\gamma^{(N)}\ ,
$$
whence the commutator of two fluctuation operators scale as a mean-field observable and becomes proportional to the identity operator within the representation fixed by the microscopic state $\omega$. 
Therefore, the fluctuation operators are expected to obey Canonical Commutation Relations 
\begin{equation}
\label{commrel}
\Big[F(s_\mu)\,,\,F(s_\nu)\Big]=i\,\epsilon_{\mu\nu\gamma}\omega(s_\gamma)\ ,
\end{equation}
and to give rise to the algebra of Weyl operators $\displaystyle W(r)=\exp\left(i(r,F)\right)$ such  that
\begin{equation}
\label{complaw}
W(r_1)W(r_2)=W(r_1+r_2)\,{\rm e}^{-\frac{i}{2}(r_1,\sigma^\omega r_2)}\ ,
\end{equation}
with $(r,F)=\sum_{\mu=1}^3r_\mu F(s_\mu)$, $r_\mu\in\mathbb{R}$, and $\sigma^\omega$ a symplectic form defined through \eqref{commrel} with entries as in \eqref{sympcov} below.

The Weyl operators are expected to emerge in the mesoscopic limit from microscopic Weyl-like spin operators of the form 
\begin{equation}
\label{locfluct}
W_N(r)=\exp\Big(i(r,F_N)\Big)\ ,\quad  (r,F_N)=\sum_{\mu=1}^3r_\mu F_N(s_\mu)\ .
\end{equation} 
The meaning of the mesoscopic limit derives from the Quantum Central Limit theorem 
that holds for states $\omega$ supporting \emph{normal quantum fluctuations} \cite{BOOK,GVVNCCL}, namely such that the quantities 
\begin{equation}
\label{sympcov}
\sigma_{\mu\nu}^\omega=-i\lim_{N\to\infty}\omega\Big(\left[F_N(s_\mu),F_N(s_\nu)\right]\Big)\ ,\qquad
\Sigma_{\mu\nu}^\omega=\frac{1}{2}\lim_{N\to\infty}\omega\Big(\left\{F_N(s_\mu),F_N(s_\nu)\right\}\Big) \ ,
\end{equation}
with $\{\cdot\,,\,\cdot\}$ denoting the anti-commutator, are well defined and give rise to a symplectic matrix $\sigma^\omega$, respectively to a covariance matrix $\Sigma^\omega$.
Then, what the \textit{Quantum Central Limit} shows is  that
\begin{equation}
\lim_{N\to\infty}\omega\left({\rm e}^{i(r,F_N)}\right)={\rm e}^{-\frac{1}{2}(r,
\Sigma^\omega r)}\ ,
\end{equation}
and similarly for products of exponentials:
$$
\lim_{N\to\infty}\omega\left({\rm e}^{i(r_1,F_N)}\,{\rm e}^{i(r_2,F_N)}\right)={\rm e}^{-\frac{1}{2}\Big(r_1+r_2,\Sigma^\omega(r_1+r_2)\Big)}\,
{\rm e}^{-\frac{i}{2}(r_1,\sigma^\omega r_2)}\ .
$$
Therefore, in the mesoscopic limit, the set of fluctuations $\{F_N(s_\mu)\}_{\mu=1}^3$ is mapped into a set of Bose field operators $\{F(s_\mu)\}_{\mu=1}^3$, so that, at the mesoscopic scale, the quantum spin chain is completely described by the algebra of Weyl operators $\displaystyle W(r)$ defined by:
\begin{equation}
\label{mes-lim}
W(r):=m-\lim_{N\to\infty}W_N(r)\ .
\end{equation}
The notation $m-\lim$ stands for a limit with respect to a topology which is weaker than the one associated with the weak limit introduced in \eqref{wlim}. This latter does indeed define a topology on the algebra of operators which is too strong to make the exponential operators $W_N(r)$ converge to definite operators.
The mesoscopic limit is understood as follows.
The quantum central limit theorems \cite{BOOK}-\cite{GVVDOF} show that, from the microscopic state $\omega$, there emerges a mesoscopic 
bosonic Gaussian state $\Omega$ on the fluctuation algebra  such that:
\begin{eqnarray}
\label{gausstate1}
\Omega\left(W(r))\right)&=&{\rm e}^{-\frac{1}{2}(r\,,\,\Sigma^\omega\, r)}\\
\nonumber
\Omega\left(W(r_1)W(r_2)\right)&=&{\rm e}^{-\frac{1}{2}((r_1+r_2)\,,\,\Sigma^\omega\,(r_1+r_2))}\,
{\rm e}^{-\frac{i}{2}(r_1,\sigma^\omega r_2)}\\
\nonumber
\Omega\left(W(r_1)W(r_2)\dots W(r_n)\right)&=&\lim_{N\to\infty}\omega\left(W_N(r_1)W_N(r_2)\dots W_N(r_n)\right)\ .
\end{eqnarray}
By varying $r_{1,2}\in\mathbb{R}^3$, the expectation values of the form $\Omega\left(W(r_1)\,X\,W(r_2)\right)$ \footnote{The expectation values $\Omega\left(W(r_1)\,X\,W(r_2)\right)$ can indeed be seen as matrix entries of $X$ with respect to vectors of a Hilbert space in the GNS-representation constructed by means of the Weyl algebra and the state $\Omega$ defined on it (see footnote \ref{foot1}).} completely determine any generic operator $X$ in the algebra generated by the the Weyl operators $W(r)$. Then, the mesoscopic convergence to $X$ of a linear combination $X_N$ of exponential operators $W_N(r)$ with $N\to\infty$ is defined by 
\begin{equation}
\label{meslim}
m-\lim_{N\to\infty}X_N=X\Leftrightarrow\lim_{N\to\infty}\omega\left(W_N(r_1)\,X_N\,W_N(r_2)\right)=\Omega\left(W(r_1)\,X\,W(r_2)\right)\qquad \forall r_{1,2}\in\mathbb{R}^3\ .
\end{equation}

\section{Mean-field dissipative dynamics}

Mean-field unitary evolutions are such that the Heisenberg dynamics of quantum observables is generated by commutators with a Hamiltonian that scales as the inverse of the number of microscopic constituents  \cite{BOOK,HT,NARN}. In analogy with these models, we here consider a simplified mean-field purely dissipative dynamics, without Hamiltonian terms, for $N$ sites corresponding, in the Heisenberg picture,  to the time-evolution equation 
\begin{equation}
\label{HME}
\partial_t x_t=\bL_N[x_t]\ ,
\end{equation}
where $\bL_N$ is a Lindblad-type generator \cite{L,GKS} of the form
\begin{equation}
\label{LindN}
\bL_N[x]=\frac{1}{N}\sum_{k,h=1}^{N}\sum_{\mu,\nu=1}^3\frac{D_{\mu\nu}}{2}\left(\left[s_\mu^{(k)}\,,\,x\right]\,s_\nu^{(h)}\,+\,s_\mu^{(k)}\,\left[x\,,\,s_\nu^{(h)}\right]\right)\ ,
\end{equation}
and $x\in\mathcal{A}$ is any spin operator supported by the sites from $1$ to $N$.
The coefficients $D_{\mu\nu}$ do not depend on the lattice sites, so one can recast the generator as follows: 
$$
\bL_N[x]=\sum_{\mu,\nu=1}^3\frac{D_{\mu\nu}}{2}\left(\left[L_\mu^{(N)}\,,\,x\right]\,
L_\nu^{(N)}\,+\,L_\mu^{(N)}\,\left[x\,,\,L_\nu^{(N)}\right]\right)\ ,\quad
L^{(N)}_\mu:=\frac{1}{\sqrt{N}}\sum_{k=1}^Ns_\mu^{(k)}\ ,
$$
with operators $L^{(N)}_\mu$ that scale as fluctuations. 
This type of generator could be derived, within the theory of open quantum system, from a suitable weak coupling \cite{AL,Petruccione} of the first $N$ spins of the chain with an environment, the interaction involving the spin-operators $L_\mu^{(N)}$.
This microscopic dissipative dynamics differs from the one studied in \cite{BCF} whose Lindblad generator is not mean-field as the operators contributing to it do not scale as fluctuations.

With the request that the $3\times 3$ Kossakowski matrix $D=[D_{\mu\nu}]$ be positive semi-definite, the generator gives rise to a dissipative semigroup consisting of completely positive, unital maps 
$\gamma_t^{(N)}=\exp(t\bL_N)$ such that $\gamma_t^{(N)}[1]=1$ for all $t\geq 0$, and
\begin{equation}
\label{semig}
\gamma^{(N)}_t\circ\gamma^{(N)}_s=\gamma_s^{(N)}\circ\gamma^{(N)}_t=\gamma^{(N)}_{t+s}\qquad\forall \ s,t\geq 0\ .
\end{equation}

Since $D=D^\dag$, then  $D^*=D^{tr}$, where $D^*$ is the matrix obtained from $D$ by taking the conjugate of all its entries and $D^{tr}$ denotes matrix transposition. Then, by decomposing $D$ as $D=A+iB$, with $A=(D+D^{tr})/2$ real symmetric and $B=(D-D^{tr})/(2i)$ real anti-symmetric, one can write $\bL_N$ as the sum of two maps: $\bL_N=\bA_N+\bB_N$, where 
\begin{eqnarray}
\label{mapA}
\bA_N[x]&=&\frac{1}{N}\sum_{k,h=1}^N\sum_{\mu,\nu=1}^3\frac{A_{\mu\nu}}{2}\left[\left[s_\mu^{(k)},x\right],s_\nu^{(h)}\right]\\
\label{mapB}
\bB_N[x]&=&\frac{i}{N}\sum_{k,h=1}^N\sum_{\mu,\nu=1}^3\frac{B_{\mu\nu}}{2}\left\{\left[s_\mu^{(k)},x\right],s_\nu^{(h)}\right\}\ .
\end{eqnarray}

As mentioned in the Introduction, we are interested in two specific scenarios: in the first one, the large $N$ limit affects only the Lindblad generator, but not the local operators $x\in\mathcal{A}$. In the second scenario, the operators $x$ will 
be taken to be collective quantum fluctuations, therefore scaling themselves as the Kraus operators.

\subsection{Mean-field dissipative dynamics of local operators}

Let us first focus on the first scenario: the large $N$ dissipative time-evolution of strictly local microscopic observables. In the large $N$ limit, the two components of the generator $\bL_N$ act very differently on local operators supported by an arbitrary but fixed number of sites.
In order to appreciate this point, consider a single spin operator $x^{(k)}$ at site $k$; since operators at different sites commute, the double commutator in \eqref{mapA} yields
\begin{equation}
\label{Ag}
\bA_N[x^{(k)}]=\frac{1}{N}\sum_{\mu,\nu=1}^3\frac{A_{\mu\nu}}{2}\left[\left[s_\mu^{(k)},x^{(k)}\right],s_\nu^{(k)}\right]\ .
\end{equation} 
The norm of $\bA_N[x^{(k)}]$ vanishes when $N\to\infty$ because of the finite number of norm-bounded contributions from the double sum. 
Instead, the anti-commutator in \eqref{mapB} gives
\begin{equation}
\label{Bg}
\bB_N[x^{(k)}]=\frac{i}{N}\sum_{\ell=1}^N\sum_{\mu,\nu=1}^3\frac{B_{\mu\nu}}{2}\left\{\left[s_\mu^{(k)},x^{(k)}\right],s_\nu^{(\ell)}\right\}\ .
\end{equation}
With respect to a clustering state $\omega$, the mean-field observable
$\displaystyle S^{(N)}_\nu=\frac{1}{N}\sum_{\ell=1}^Ns_\nu^{(\ell)}$ tends to a scalar quantity $\omega_\nu:=\omega(s_\nu)$ 
and thus $\bB_N[x^{(k)}]$ tends weakly to a state-dependent  Hamiltonian action 
$$
w-\lim_{N\to\infty}\bB_N[x^{(k)}]=i\left[H^{(k)}_\omega\,,\,x^{(k)}\right]\ ,\qquad H^{(k)}_\omega=\sum_{\mu,\nu=1}^3B_{\mu\nu}\,\omega_\nu\,s_\mu^{(k)}\ .
$$
Since the entries $B_{\mu\nu}$ are real as well as the expectations $\omega_\nu$ and $s_\nu=s_\nu^\dag$, $H^{(k)}_\omega$ is Hermitean.

Furthermore, looking more accurately at the time-evolution equation \eqref{HME}, one sees that $\bL_N$ acts on $x_t=\gamma^{(N)}_t[x]$ so that one needs to study the large $N$ behaviour of  
$$
\bL_N[x_t]=\gamma^{(N)}_t\left[\bL_N[x]\right]=\gamma^{(N)}_t\left[\bA_N[x]\right]+\gamma^{(N)}_t\left[\bB_N[x]\right]\ .
$$
If  $\omega^{(N)}_t:=\omega\circ\gamma^{(N)}_t$ tends on local spin operators $x\in\mathcal{A}$ to 
a still clustering, but possibly time-dependent state,
\begin{equation}
\label{asymstate}
\omega_t(x)=\lim_{N\to\infty}\omega^{(N)}_t(x)\ ,
\end{equation}
then the emergent Hamiltonian will also be time-dependent,
\begin{equation}
\label{tHam}
w-\lim_{N\to\infty}\gamma^{(N)}_t\left[\bB_N[x]\right]=i\sum_{k}\left[H^{(k)}_{\omega_t}\,,\,x\right]\ ,\qquad H^{(k)}_{\omega_t}=\sum_{\mu,\nu=1}^3B_{\mu\nu}\,\omega_\nu(t)\,s_\mu^{(k)}\ ,
\end{equation}
due the time-dependent macroscopic averages 
\begin{equation}
\label{macrobs}
\omega_\nu(t)=\lim_{N\to\infty}\omega^{(N)}_t\left(S^{(N)}_\nu\right)=\lim_{N\to\infty}\omega\left(\gamma^{(N)}_t[S^{(N)}_\nu]\right)\ .
\end{equation}

Starting from an arbitrary initial time $t_0\geq 0$, the emergent unitary dynamics on local observables $x\in\mathcal{A}$ will thus amount to a homogenous one-parameter family of maps  $\alpha^\omega_{t-t_0}$ such that
\begin{equation}
w-\lim_{N\to\infty}{\rm e}^{(t-t_0)\bL_N}[x]=\alpha^\omega_{t-t_0}[x]\ .
\label{homo}
\end{equation}
It thus follows that, in the large $N$ limit, the irreversible, purely dissipative semigroup of maps $\gamma^{(N)}_t$, not only acts on local observables as a family of unitary maps $\alpha_t^\omega$, but also that these break the microscopic composition law \eqref{semig}. Indeed, the generator of these maps depends explicitly not only on the final time $t$, but also on the initial time $t_0$:
\begin{equation}
\frac{d}{dt}\alpha^\omega_{t-t_0}[x^{(k)}]=i\sum_{\mu,\nu=1}^{3}B_{\mu\nu}\,\omega_\nu(t-t_0)\,\left[\alpha^\omega_{t-t_0}\left(s_\mu^{(k)}\right)\,,\,\alpha^\omega_{t-t_0}\left(x^{(k)}\right)\right]\ .
\label{nonloc}
\end{equation}

\begin{remark}
\label{rem1}
For each initial time $t_0\geq 0$ we have a one-parameter family of maps
$\alpha^\omega_{t-t_0}$ that obey neither the composition law in \eqref{semig}, 
nor the one typical of two-parameter semi-groups, 
$$
\alpha_{t,t_0}=\alpha_{t,s}\circ\alpha_{s,t_0}\qquad\forall 0\leq t_0\leq s\leq t\ ,
$$
that arises from time-ordered integration of a generator depending explicitly on the running time $t$, but not on the initial time $t_0$. 
The Hamiltonian in \eqref{nonloc} thus provides an interesting instance of non-Markovianity in the sense of \cite{ChrKos}. 
Clearly, if the time-evolving state $\omega\circ\gamma^{(N)}_t$ tends, with $N\to\infty$ to a time-invariant state on the quasi-local algebra $\mathcal{A}$, then one recovers the one-parameter semigroup features of \eqref{semig}. Indeed, the time-evolution is in this case also Hamiltonian, but obtained from a micro-dynamics of semigroup type; therefore, the obtained time-evolution of local operators also holds only for positive times. 
\end{remark}

\subsection{Mean-field dissipative dynamics of quantum fluctuations}

Suppose the state $\omega$ is not left invariant by the microscopic dynamics so that $\omega^{(N)}_t:=\omega\circ\gamma^{(N)}_t\neq\omega$ and that the mean-values in \eqref{fluc} are time-dependent. Then, as quantum fluctuations account for deviations 
from mean-values evaluated with respect to the chosen reference state that now varies in time, they are themselves time-dependent and defined by
\begin{equation}
\label{timefluct}
F_t^{(N)}(s_\mu)=\frac{1}{\sqrt{N}}\sum_{k=1}^N\left(s_\mu^{(k)}-\omega^{(N)}_t(s_\mu^{(k)})\right) \ .
\end{equation}
Notice that 
\begin{equation}
\label{timefluct2}
\omega^{(N)}_t(F^{(N)}_t(s_\mu))=0\ .
\end{equation}

If the macroscopic triples $(\omega_1(t),\omega_2(t),\omega_3(t))$, where 
$\omega_\nu(t)=\lim_{N\to\infty}\omega^{(N)}_t(S^{(N)}_\mu)$, are time-dependent, then one must
deal with time-dependent symplectic and covariance matrices,  
$\sigma^\omega_t$ and $\Sigma^\omega_t$, with entries (compare them with \eqref{sympcov})
\begin{eqnarray}
\label{Sympt}
\sigma^\omega_{\mu\nu}(t)&=&-i\lim_{N\to\infty}\omega^{(N)}_t\Big(\left[F_t^{(N)}(s_\mu)\,,\,F_t^{(N)}(s_\nu)\right]\Big)\ ,\\
\Sigma^\omega_{\mu\nu}(t)&=&\frac{1}{2}\lim_{N\to\infty}\omega^{(N)}_t\Big(\left\{F_t^{(N)}(s_\mu)\,,\,F_t^{(N)}(s_\nu)\right\}\Big)\ .
\label{covt}
\end{eqnarray}
In such a case, the dissipative dynamics of quantum fluctuations consists of a one-parameter family of completely positive maps turning Weyl operators into themselves in the sense that \cite{BCFN}:
\begin{eqnarray}
\lim_{N\to\infty}\omega^{(N)}_t\left({\rm e}^{i(r,F^{(N)}_t)}\right)
&=&\Omega_t\left(W(r)\right)={\rm e}^{-1/2(r,\Sigma^\omega_t r)}\ ,
\label{dynflu1}\\
\label{dynflu2}
\Omega_t\left(W(r_1)W(r_2)\right)&=&{\rm e}^{-\frac{1}{2}\big((r_1+r_2)\,,\,\Sigma^\omega_t\,(r_1+r_2)\big)}\,
{\rm e}^{-\frac{i}{2}(r_1,\sigma^\omega_t r_2)}\ ,
\end{eqnarray}
where $\Sigma^\omega_t=X_t\,\Sigma^\omega\,X_t^{tr}\,+\,Y_t$ with
$X_t$ and $Y_t\geq 0$ being fairly complicated matrices \cite{BCFN} reflecting the dependence of the time-evolution on the evolving quasi-local state. Quantum fluctuations thus evolve dissipatively and in a non-Markovian fashion as is the case for the unitary dynamics of local observables. More precisely, starting at an initial time $t_0\geq0$, the dynamics of quantum fluctuations arises from the following mesoscopic limit (see \eqref{mes-lim})
$$
\lim_{N\to\infty}\omega\left(W_N(r_1)\gamma^{(N)}_{t-t_0}\left[{\rm e}^{i(r,F^{(N)}_{t-t_0}}]\right]\,W_N(r_2)\right)
=\Omega\left(W(r_1)\,\Phi_{t-t_0}[W(r)]\,W(r_2)\right)\ \quad \forall r,r_{1,2}\in\mathbb{R}^3\ .
$$
The maps $\Phi_{t-t_0}$ can then be shown to be unital, completely positive by applying the techniques developed in \cite{CPCCR} and also to be such that $\Phi_{t-t_0}\neq\Phi_{t-s}\circ\Phi_{s-t_0}$ for $t_0\leq s\leq t$ (see \cite{BCFN}).

The matrices $X_t$ and $Y_t$ result from the integration of the matrix equation satisfied by the time-derivative of the covariance matrix $\Sigma^\omega_t$:
\begin{equation}
\label{Smateq}
\dot{\Sigma}^\omega_t=\sigma^\omega_t\,A\,(\sigma_t^\omega)^{tr}\,+\,\Big(\sigma^\omega_t\,B\,\Sigma^\omega_t\,+\,\Sigma^\omega_t\,B\,\sigma^\omega_t\Big)\,+\,\Big(C_t\,\Sigma^\omega_t\,+\,\Sigma^\omega_t\,C^{tr}_t\Big)\ .
\end{equation}
whose derivation is sketched in the Appendix, the matrix $A$ and $B$ being those in \eqref{mapA} and \eqref{mapB}, while the matrix $C_t$ is given in equation \eqref{Bg5b} of the Appendix.

\begin{remark}
\label{rem2}
If in  the left hand side of \eqref{dynflu2} one uses the composition law \eqref{complaw} and the linearity of the expectation given by the state $\Omega_t$ one derives $\sigma^\omega_t=\sigma^\omega$, a contradiction when the macroscopic triple $(\omega_1(t),\omega_2(t),\omega_3(t))$ depends on time.
It thus follows that when evaluating the averages at time $t$ of mesoscopic Weyl operators multiplied by functions $f(\mathbf{\cdot})$ of the triple $\vec{\omega}=(\omega_1,\omega_2,\omega_3)$, $\omega_\mu=\omega(s_\mu)\in[-1/2,1/2]$, these functions ought to be evaluated at the time-evolved macroscopic triple $\vec{\omega}_t=(\omega_1(t),\omega_2(t),\omega_3(t))$:
$$
\Omega_t\Big(W(r)f(\cdot)\Big)=\Omega_t(W(r))\,f(\vec{\omega}_t)\ .
$$
When $f(\vec{\omega})=\exp(-i(r,\sigma^\omega r))$, the contradiction mentioned above is eliminated. 
We refer the reader to \cite{BCFN} for a detailed discussion of this  point.
\end{remark}
\medskip

\begin{remark}
\label{rem3}
As discussed in the previous section, on the quasi-local spin algebra $\mathcal{A}$ the dissipative time-evolution $\gamma_t^{(N)}$ behaves, in the large $N$ limit, as the unitary map $\alpha^\omega_t$ when $N\to\infty$. There thus arises the possibility of constructing the mesoscopic description based on the, in general time-dependent, microscopic state $\hat{\omega}_t:=\omega\circ\alpha^\omega_t$ and on the corresponding fluctuations
\begin{equation}
\label{dynflu3}
\hat{F}_t^{(N)}(s_\mu)=\frac{1}{\sqrt{N}}\sum_{k=1}^N\Big(s^{(k)}_\mu-\hat{\omega}_t
(s^{(k)}_\mu)\Big)\ .
\end{equation}
The emerging mesoscopic state $\hat{\Omega}_t$ such that
\begin{equation}
\label{qltQCL}
\lim_{N\to\infty}\omega\circ \alpha_{t}^\omega\left({\rm e}^{i(r,F_t^{(N)})}\right)=\hat{\Omega}_t\left(W(r)\right)={\rm e}^{-1/2(r,\hat{\Sigma}_t r)}\ ,
\end{equation}
and the mesoscopic dynamics will in general differ from those associated with the states
$\omega^{(N)}_t$ as we shall show in Section \ref{mescorr} (for similar considerations in the case of unitary time-evolutions see \cite{HT,NARN}).
\end{remark}

\section{Asymptotic invariant state}

As we have seen, the large $N$ dynamics of both local operators and quantum fluctuations
depends on the time-dependence of 
the macroscopic averages $\omega_\nu(t)$ in \eqref{macrobs}.
By taking the time-derivative of \eqref{macrobs} and using \eqref{tHam}, the real anti-symmetric character of the matrix $B=[B_{\mu\nu}]$ yields the following non-linear differential system:
\begin{eqnarray}
\label{eq1}
\dot{\omega}_1(t)&=&-B_{12}\,\omega_1(t)\,\omega_3(t)\,+\,B_{13}\,\omega_1(t)\,\omega_2(t)\,+\,B_{23}\,(\omega_2^2(t) +\omega_3^2(t))\ ,\\
\label{eq2}
\dot{\omega}_2(t)&=&-B_{12}\,\omega_2(t)\,\omega_3(t)\,-B_{23}\,\omega_1(t)\,\omega_2(t)\,-B_{13}\,(\omega_1^2(t) +\omega_3^2(t))\ ,\\
\label{eq3}
\dot{\omega}_3(t)&=&B_{13}\,\omega_2(t)\,\omega_3(t)\,-\,B_{23}\,\omega_1(t)\,\omega_3(t)\,+\,B_{12}\,(\omega_1^2(t)+\omega_2^2(t))\ .
\end{eqnarray}
By a suitable unitary rotation that preserves the algebraic relations between the fluctuations $F_N(s_\mu)$, $B$ can always be brought into the simpler form:
\begin{equation}
\label{simpleB}
B=\begin{pmatrix}
0&\lambda&0\\
-\lambda&0&0\\
0&0&0
\end{pmatrix}\ .
\end{equation}
Correspondingly, the spin operators $s_\mu$ rotate into new spin operators 
$s'_\mu$ and the mean-values $\omega_\mu(t)$ into new mean-values $\omega'_\mu(t)$.
For sake of simplicity, we shall denote the new spin operators and their mean-values as the old ones; then, from \eqref{simpleB} it follows that 
$$
\dot{\omega}_1(t)=-\lambda\, \omega_1(t)\,\omega_3(t)\ ,\quad
\dot{\omega}_2(t)=-\lambda\, \omega_2(t)\,\omega_3(t)\ ,\quad
\dot{\omega}_3(t)=\lambda\,(\omega_1^2(t) +\omega_2^2(t))\ .
$$
One then readily sees that the length of the vector with components $\omega_\nu(t)$ 
remains constant under the given unitary time-evolution; then, setting $\omega_1^2(t)+\omega_2^2(t)+\omega_3^2(t)=\xi^2$, the last equation reads 
$\displaystyle
\dot{\omega}_3(t)=\lambda(\xi^2-\omega_3^2(t))$ so that
$$
\omega_3(t)=\xi\tanh(\xi(\lambda t+c))\ ,\quad  
\omega_{1,2}(t)=\frac{\cosh(\xi c)}{\cosh(\xi(\lambda t+c))}\,\omega_{1,2}(0)\ ,
$$
where the constant $c$ is determined by the initial condition through 
$\omega_3(0)=\xi\tanh(\xi c)$.
It thus follows that the only time-invariant solutions are:
\begin{equation}
\label{asympt2}
\omega_1(\infty)=\omega_2(\infty)=0,\qquad \omega_3(\infty)=\xi=\pm |\xi|\ .
\end{equation}
For $\lambda>0$, $\omega_3(\infty)=|\xi|$ is a stable solution while $\omega_3(\infty)=-|\xi|$ an unstable one; the opposite being true for $\lambda<0$; whence, 
starting from any initial triple $(\omega_1,\omega_2,\omega_3)$, except from $(0,0,\xi)$ such that $\lambda\xi<0$, one always converges to the stable solution 
$(0,0,\xi)$.

\subsection{Time-invariant macroscopic averages}

We now study the emergent quantum fluctuation dynamics when the spin chain is equipped with a time-invariant microscopic state $\omega$ such that the macroscopic averages 
$\omega_\mu(t)$  are constant in time and equal the asymptotic values $(0,0,\xi)$ discussed above \footnote{The simplest example of a microscopic state $\omega$ with such properties is the (infinite) tensor product of a same density matrix $\rho$ such that at each site ${\rm Tr}(\rho\,s_{1,2})=0$ and ${\rm Tr}(\rho\,s_3)=\xi$.\label{foot2}}.

With respect to such a microscopic state, the algebra of quantum fluctuations reduces to the Weyl algebra acting on the one-particle Hilbert space $L^2(\mathbb{R})$.  
In fact, recalling the definition of quantum fluctuations given in \eqref{fluc} and their
commutation relations \eqref{commrel}, one gets:
\begin{equation}
[F( s_1),F( s_2)]=i\, \xi,\qquad [F(s_3),F(s_{1,2,3})]=0\ .
\end{equation}
Therefore, $F(s_{1,2})$ behave as position and momentum operators
\begin{equation}
q=\frac{1}{\sqrt{|\xi|}}F( s_1),\qquad p=\frac{\sgn(\xi)}{\sqrt{|\xi|}}F(s_2)\ ,\qquad
[q,p]=i \ ,
\label{toxp}
\end{equation}
while $F( s_3)$ corresponds to a classical degree of freedom commuting with both $q$ and $p$.

Let us now consider the two maps in \eqref{mapA} and \eqref{mapB}, where the spin operators
that define them are those relative to the representation \eqref{simpleB} of the matrix 
$B$ and, in the 
place of $x_t$, there is an exponential operator $W_N(r)$ as in \eqref{locfluct}. In the mesoscopic limit \eqref{mes-lim}, $W_N(r)$ tends to a Weil operator $W(r)$ and the map \eqref{mapA} defined by the real part $A$ of the Kossakowwski matrix $D$ gives rise to
\begin{eqnarray}
\label{flucinvA1}
\bA[W(r)]&=&\sum_{\mu,\nu=1}^3\frac{A_{\mu\nu}}{2}\Big[\Big[F(s_\mu)\,,\,W(r)\Big]\,,\,F(s_\nu)\Big]\\
\label{flucinvA2}
&=&\frac{|\xi|}{2}\,\Big(A_{11}\,\Big[\Big[q\,,\,W(r)\Big]\,,\,q\Big]\,+\,
A_{22}\Big[\Big[p\,,\,W(r)\Big]\,,\,p\Big]\Big)\\
\label{flucinvA3}
&+&\frac{\xi}{2}\Big(A_{12}\Big[\Big[q\,,\,W(r)\Big]\,,\,p\Big]\,+\,A_{21}\Big[\Big[p\,,\,W(r)\Big]\,,\,q\Big]\Big)\ .
\end{eqnarray}
Indeed, the double commutator in \eqref{Ag} allows one to replace the sums $\sum_{k=1}^Ns^{(k)}_\mu/\sqrt{N}$ with fluctuations operators $F_N(s_\mu)$ by inserting the needed mean-values; then, the expressions \eqref{flucinvA2} and \eqref{flucinvA3} in terms of $q,p$ follow from \eqref{toxp}.

Notice that, since $F(s_3)$ commutes with $F(s_{1,2})$, the Weyl operators can be split as 
\begin{equation}
\label{Weylsplit}
W(r)=\exp(i(r_1 F(s_1)+r_2F(s_2))\,\exp(ir_3F(s_3))\ ,
\end{equation}
whence one can restrict to Weyl operators involving only $F(s_{1,2})$ or, equivalently $q$ and $p$.

Regarding the map \eqref{mapB} defined by the imaginary part $B$ of the Kossakowski matrix, only spin operators $s^{(k)}_{1,2}$ appear and they have vanishing  mean-values with respect to $\omega^{(N)}_t$ when $N\to\infty$; indeed, since $\omega$ is translation invariant and gives rise to the invariant macroscopic triple $(0,0,\xi)$, $\alpha^\omega_t$ acts trivially so that
\begin{equation}
\label{invas}
\lim_{N\to\infty}\omega^{(N)}_t(s^{(k)}_{1,2})=\omega\circ\alpha^\omega_t(s^{(k)}_{1,2})
=\omega(s_{1,2}^{(k)})=\omega(s_{1,2})=0\ .
\end{equation}

Then, the action of $\bB_N[W_N(r)]$ can be readily expressed in terms of the fluctuation operators $F_N(s_{1,2})$ whence the mesoscopic limit yields
\begin{eqnarray}
\label{flucinv1}
\bB[W(r)]&=&\frac{i\lambda}{2}\Big(\Big\{\Big[F( s_1)\,,\,W(r)\Big]\,,\, F( s_2)\Big\}-\Big\{\Big[F( s_2)\,,\,W(r)\Big]\,,\,F(s_1)\Big\}\Big)\\
\label{flucinv2}
&=&\frac{ib}{2}\Big(\Big\{\Big[q\,,\,W(r)\Big]\,,\,p\Big\}-\Big\{\Big[p\,,\,W(r)\Big]\,,\,q\Big\}\Big)\ ,
\end{eqnarray}
where $b=\lambda\,\xi$.  
As a consequence, the stability of the macroscopic triple $(0,0,\xi)$ amounts to $b>0$, its instability to  $b<0$.
The dynamics thus keeps the bosonic mode $F(s_3)$ commuting with $q$ and $p$ at all times so that, using \eqref{Weylsplit}, we can concentrate on Weyl operators of the form $W(r)=\exp(i(r_1 q+r_2p))$. Their mesoscopic dynamics is thus generated by
\begin{equation}
\bL[W(r)]=\sum_{\mu,\nu=1}^{2}\frac{D_{\mu\nu}}{2}\Big(\Big[R_\mu\,,\,W(r)\Big]\,R_\nu\,+\,R_\mu\,\Big[W(r)\,,\,R_\nu\Big]\Big)\ ,
\label{gengen}
\end{equation}
with $R=(q,p)^{tr}$ and Kossakowski matrix given by
\begin{equation}
\label{newK}
D=A\,+\,i\,B\ ,\quad A=\begin{pmatrix}
A_{11}\,|\xi|&A_{12}\,\xi\\
A_{12}\,\xi&A_{22}\,|\xi|
\end{pmatrix}\ ,\qquad B=\begin{pmatrix}
0&b\\
-b&0
\end{pmatrix}\ .
\end{equation}
Its action on  $W(r)=\exp(i(r,R))$ can be unfolded by using that 
$$
W^\dag(r)\,q\,W(r)=q\,-\,r_2\ ,\qquad W^\dag(r)\,p\,W(r)=p\,+\,r_1\ .
$$
Then, 
$$
\bL[W(r)]=\left(-i\,b\,(r\,,\,R)\,+\,\frac{1}{2}\left(r,\sigma^\omega\,A\,\sigma^\omega\,r\right)\right)\,W(r)\ ,\qquad  \sigma^\omega=\begin{pmatrix}
0&1\\
-1&0
\end{pmatrix}\ .
$$
On the other hand, since $\bL$ is bilinear in $q$ and $p$, the dynamics 
$\exp(t\bL)$ transform Weyl operators into Weyl operators multiplied by scalar functions.
Therefore, with the ansatz
\begin{equation}
\label{fluctev}
W_t(r)={\rm e}^{t\bL}\left[W(r)\right]=\exp\left(i(r_t\,,\,R)\right)\,\exp\left(
-\frac{1}{2}(r,Y_t\,r)\right)\ ,
\end{equation}
from $\partial_tW_t(r)=\bL\left[W_t(r)\right]$ and
$$
\partial_tW_t(r)=\left(i(\dot{r}_t\,,\,R)\,+\,\frac{1}{2}\Big(
i(\dot{r}_t\,,\,\sigma^\omega\,r_t)-(r,\dot{Y}_t\,r)\Big)\right)\,W_t(r)\ ,
$$
one then gets
\begin{equation}
\label{eqns}
r_t={\rm e}^{-b\,t}\,r\ ,\qquad
\dot{Y}_t=-{\rm e}^{-2b\,t}\,\sigma^\omega\,A\,\sigma^\omega\ ,\qquad Y_0=0\ ;
\end{equation}
indeed, $(\dot{r}_t\,,\,\sigma^\omega\,r_t)=-b(r_t\,,\,\sigma^\omega\,r_t)=0$ due to the anti-symmetric character of $\sigma^\omega$.

Any initial Gaussian mesoscopic state $\Omega^G$ is thus mapped, at time $t>0$, into a Gaussian state $\Omega^G_t$ completely defined by its covariance matrix $\Sigma_t$ with entries
$\frac{1}{2}\Omega^G_t\left(\left\{R_\mu,R_\nu\right\}\right)$:
\begin{equation}
\Sigma^G_t=e^{-2bt}\begin{pmatrix}
\Sigma^G_{11}&\Sigma^G_{12}\\
\Sigma^G_{12}&\Sigma^G_{22}
\end{pmatrix}
-\frac{1-{\rm e}^{-2bt}}{2b}\,\sigma^\omega\,A\,\sigma^\omega\ .
\label{covflu}
\end{equation}
\medskip

\noindent
$\mathbf{b>0}\,$:
the macroscopic triple $(0,0,\omega_3(\infty)=\xi)$ is stable; then, letting $t\to+\infty$, the mesoscopic state $\Omega^G_t$ tends to an asymptotic 
time-invariant Gaussian state $\Omega_\infty$ with covariance matrix
\begin{equation}
\Sigma_\infty:=\lim_{t\to+\infty}\Sigma^G_t=\frac{1}{2b}\begin{pmatrix}
|\xi|\,A_{22}&-A_{12}\,\xi\\
-\xi\,A_{12}&|\xi|\,A_{11}
\end{pmatrix}\ .
\label{asycovflu}
\end{equation}
\medskip

\noindent
$\mathbf{b<0}$\,: 
the macroscopic time-invariant triple $(0,0,\omega_3(\infty)=\xi)$ is unstable; the covariance matrix at time $t$ reads
\begin{equation}
\Sigma^G_t={\rm e}^{2|b|t}\begin{pmatrix}
\Sigma^G_{11}&\Sigma^G_{12}\\
\Sigma^G_{12}&\Sigma^G_{22}
\end{pmatrix}\,+\,\frac{1}{2|b|}\begin{pmatrix}
|\xi|\,A_{22}&-\xi\, A_{12}\\
-\xi\,A_{12}&|\xi|\, A_{11}
\end{pmatrix}\,\left({\rm e}^{2|b|t}-1\right)\ .
\end{equation}
Both matrices contributing to $\Sigma^G_t$ are positive semi-definite; therefore, their diagonal entries are non-negative, whence 
the norm of $\Sigma^G_t$ grows exponentially with  $t\to+\infty$, reflecting the instability of the invariant macroscopic triple. Therefore, in the unstable case, no invariant mesoscopic Gaussian states can exist.

\begin{example}
Let us consider a generator as in \eqref{gengen} with the following 
Kossakowski matrix
\begin{equation}
\label{simpleD}
D=\begin{pmatrix}
1&ib\\
-ib&b^2
\end{pmatrix}\ .
\end{equation}
In the Schr\"odinger picture, mesoscopic states (density matrices) $\rho$ on the Weyl algebra evolve in time according to the master equation $\partial_t\rho_t=\bL^T[\rho_t]$, where 
$$
\bL^T[\rho_t]=\left(q+ibp\right)\rho_t\left(q-ibp\right)-\frac{1}{2}\Big\{\left(q-i\,b\,p\right)\left(q+i\,b\,p\right),\rho_t\Big\}\ .
$$

\noindent
\textbf{Stable case: $b=1$}.\\
One can represent the action of $q+i\,b\,p=q+ip$ by means of an annihilation operator $a$ and
seek invariant states such that $\bL^T[\rho]=0$ by expanding $\rho=\sum_{n,m\geq 0}\rho_{nm}\vert n\rangle\langle m\vert$, with $a|n\rangle=\sqrt{n}\vert n-1\rangle$.
Imposing 
\begin{eqnarray*}
0&=&a\sum_{n,m\geq 0}\rho_{nm}\,\vert n\rangle\langle m\vert a^\dagger-\frac{1}{2}a^{\dagger}a\sum_{n,m}\rho_{nm}\vert n\rangle\langle m\vert\,-\,\frac{1}{2}\sum_{n,m}\rho_{nm}\vert n\rangle\langle m\vert a^{\dagger}a\\
&=&\sum_{n,m\geq 0}\left(\rho_{n+1m+1}\sqrt{(n+1)\,(m+1)}\,-\,\frac{n+m}{2}\rho_{nm}\right)\vert n\rangle\langle m\vert\ ,
\end{eqnarray*}
one derives the following recursion relation
$$
\rho_{n+1n+1}=\frac{n}{n+1}\,\rho_{nn}\qquad\forall n\geq 0\ .
$$
Then, ${\rm Tr}(\rho)=1=\sum_{n\geq 0}\rho_{nn}$ requires $\rho_{00}=1$ whence the vacuum state
$\rho=|0\rangle\langle0|$ is the only stationary state. With respect to the operators $q,p$ the vacuum state is a Gaussian state $\Omega$ with covariance matrix $\Sigma=1/2$.
Since, with $b=1$, $A$ in \eqref{newK} is the identity matrix, the vacuum state coincides with the limiting state $\Omega_\infty$ with covariance matrix $\Sigma_\infty$ in \eqref{asycovflu}.
\medskip

\noindent
\textbf{Unstable case: $b=-1$}\\
By representing the action of $q+i\,b\,p=q-i\,p$ with the annihilation operator $a$, invariant states must satisfy:
\begin{eqnarray*}
0&=&a^\dagger\sum_{n,m\geq0}\rho_{nm}|n\rangle\langle m|a-\frac{1}{2}aa^{\dagger}\sum_{n,m\geq0}\rho_{nm}|n\rangle\langle m|-\frac{1}{2}\sum_{n,m\geq 0}\rho_{nm}|n\rangle\langle m|aa^{\dagger}\\
&=&\sum_{n,m\geq 1}\left(\rho_{n-1m-1}\sqrt{n\,m)}\,-\,\frac{2+n+m}{2}\rho_{nm}\right)\vert n\rangle\langle m\vert\,-\,\rho_{00}\vert 0\rangle\langle 0\vert\\
&-&\sum_{n\geq 1}\frac{2+n}{2}\Big(\rho_{0n}\,\vert 0\rangle \langle n\vert+\rho_{n0}\,\vert n\rangle\langle 0\vert\Big)\ .
\end{eqnarray*}
In this case there are no solutions since, unlike before, $\rho_{00}=0$;
therefore, no mesoscopic invariant states exist.
\end{example}

\subsection{Time-dependent macroscopic averages}

If the microscopic state on the quantum spin chain provides a time-dependent macroscopic triple $(\omega_1(t),\omega_2(t),\omega_3(t))$, the algebra of quantum fluctuations consists of three bosonic degrees of freedom satisfying time-dependent canonical commutation relations as in \eqref{Sympt}.
The mesoscopic dynamics $\Omega^G\to\Omega^G_t$ of an initial mesoscopic Gaussian state 
$\Omega^G$ is then completely defined by the time-evolution equation \eqref{Smateq}
satisfied by its covariance matrix $\Sigma^G_t$. The time behaviour of the latter is 
in turn determined by the dynamics of macroscopic averages, through the symplectic matrix 
\begin{equation}
\label{symplambda}
\sigma^\omega_t=\begin{pmatrix}0&\omega_3(t)&-\omega_2(t)\cr
-\omega_3(t)&0&\omega_1(t)\cr
\omega_2(t)&-\omega_1(t)&0
\end{pmatrix}\ ,
\end{equation}
and the matrix $C_t$ in \eqref{Smateq} which, with the choice of $B$ in \eqref{simpleB}, reads
\begin{equation}
\label{matD}
C_t=-\lambda\begin{pmatrix}0&0&\omega_1(t)\cr
0&0&\omega_2(t)\cr
-\omega_1(t)&-\omega_2(t)&0
\end{pmatrix}\ .
\end{equation}

If the microscopic state does not provide the unstable macroscopic triple $(0,0,\xi)$ with $b=\lambda\xi<0$, then any initial triple $(\omega_1,\omega_2,\omega_3)$ tends exponentially fast to $(0,0,\xi)$ with $b>0$.
Therefore, for a generic (translation-invariant and clustering) microscopic state, the corresponding mesoscopic symplectic matrix $\sigma^\omega_t$ will tend to
\begin{equation}
\label{sympas}
\sigma^\omega_\infty=\xi\begin{pmatrix}0&1&0\cr
-1&0&0\cr
0&0&0
\end{pmatrix}
\end{equation}
with $t\to+\infty$, while $D_t$ will vanish.
Therefore, for large times, the covariance matrix satisfies the following asymptotic version of \eqref{Smateq} with $\sigma_\infty^\omega$ as in \eqref{sympas}:
\begin{eqnarray}
\nonumber
\dot{\Sigma}^G_t&\simeq& -\,b\,\left(\begin{pmatrix}1&0&0\cr
0&1&0\cr
0&0&0\end{pmatrix}\,\Sigma^G_t\,+\,\Sigma^G_t\,\begin{pmatrix}1&0&0\cr
0&1&0\cr
0&0&0\end{pmatrix}\right)\,-\,\sigma^\omega_\infty\,A\,\sigma^\omega_\infty\\
\nonumber
\\
\label{asympteq}
&=&
-b\begin{pmatrix}2\Sigma^G_{11}(t)&2\Sigma^G_{12}(t)&\Sigma^G_{13}(t)\cr
2\Sigma^G_{12}(t)&2\Sigma^G_{22}(t)&\Sigma^G_{23}(t)\cr
\Sigma^G_{13}(t)&\Sigma^G_{23}(t)&0\end{pmatrix}\,+\,\begin{pmatrix}A_{22}&-A_{12}&0\cr
-A_{12}&A_{11}&0\cr
0&0&0\end{pmatrix}\ .
\end{eqnarray}
From such an asymptotic equation one derives that the covariance matrix $\Sigma_t^{(2)}$ relative to the mesoscopic modes $F(s_{1,2})$ asymptotically behaves as the solution
of
$$
\dot{\Sigma}^{(2)}_t=-2\,b\,\Sigma^{(2)}_t\,+\,A^{(2)}\ ,\quad A^{(2)}=\begin{pmatrix}A_{22}&-A_{12}\cr
-A_{12}&A_{11}\end{pmatrix}\ ,
$$
namely 
$\displaystyle
\Sigma^{(2)}_t={\rm e}^{-2\,b\,t}\Sigma^{(2)}\,+\,\frac{1}{2\,b}\,A^{(2)}$,
which then tends to the asymptotic covariance matrix $\Sigma_\infty$ in \eqref{asycovflu}. This shows that, in the stable regime, the resulting mesoscopic limit state is the one described by the covariance matrix in \eqref{asycovflu} as one can see by passing from $F(s_{1,2})$ to position and momentum operators. 
Namely, when restricted to the Weyl algebra generated by $F(s_{1,2})$ the time-evolving state $\Omega_t$ is such that
$\lim_{t\to+\infty} \Omega_t=\Omega_\infty$.

Different is the asymptotic behaviour of the third bosonic degree of freedom $F(s_3)$.
This cannot be extracted from the asymptotic equation \eqref{asympteq}, as it just says that asymptotically $\Sigma_{33}$ goes to a constant. Its value has to be calculated 
from the explicit solution of \eqref{Smateq}. What one finds is that, asymptotically, $F(s_3)$ gets dynamically decoupled from $F(s_{1,2})$ and that its mean-value with respect to $\Omega_t$ tends to the following Gaussian distribution:
\begin{equation}
\label{deg3}
\lim_{t\to+\infty}\Omega_t\left(e^{ir_3 F(s_3)}\right)=e^{-\frac{r_3^2}{2}s^2}\ ,\quad
s^2=\frac{1}{|\xi|^2}\sum_{\mu,\nu=1}^3\omega_i(0)\omega_j(0)\Sigma^\omega_{\mu\nu}\ .
\end{equation}
Thus, the asymptotic distribution carries memory of the initial microscopic state through the macroscopic mean-values $\omega_\nu(0)$, and of the initial mesoscopic state through the entries of the covariance matrix $\Sigma^\omega$. The dependence on theses initial states is instead lost in the quasi-local limiting state and in the mesoscopic covariance matrix \eqref{asycovflu}.

\subsection{Dynamics of mesoscopic correlations}
\label{mescorr}
    
Let $\omega$ be a stable factor state, namely the tensor product of infinitely many copies of a same density matrix for each lattice site (see Footnote \ref{foot2}), and consider the matrix $B$ in the form \eqref{simpleB}.
Then, all the Hamiltonian contributions to the local dynamics in \eqref{tHam} vanish and 
the autmorphisms $\alpha^\omega_t$ act trivially on the quasi-local algebra $\mathcal{A}$.

As emphasised in Remark \ref{rem3}, one can look at the mesoscopic dynamics that emerges from fluctuations built with respect to the time-varying state $\hat{\omega}_t=\omega\circ\alpha_t^\omega$. Indeed, one knows that on local observables $\hat{\omega}_t=\lim_N\omega^{(N)}_t$, where $\omega^{(N)}_t=\omega\circ\gamma^{(N)}_t$ is the microscopic state time-evolution under the dissipative microscopic dynamics.

In the case we are discussing we thus have that $\hat{\omega}_t=\omega$; therefore, the
fluctuation operators in \eqref{dynflu3} now read:
$$
F_N(s_{1,2})=\frac{1}{\sqrt{N}}\sum_{k=1}^Ns^{(k)}_{1,2}\ ,\qquad
F_N(s_3)=\frac{1}{\sqrt{N}}\sum_{k=1}^N\Big(s^{(k)}_3-\xi\Big)\ ,
$$
their symplectic matrix is as in \eqref{sympas}, while the covariance matrix $\Sigma^\omega$ is given by 
\begin{equation}
\label{covhat}
\Sigma^\omega=\frac{1}{4}\begin{pmatrix}
1&0&0\cr
0&1&0\cr
0&0&1-4\xi^2
\end{pmatrix}\ .
\end{equation}
This form corresponds to the fact that $\omega$ carry no correlations between any pairs of $s^{(i)}_{1}$, $s^{(j)}_{2}$ at different sites $i\neq j$.
Notice that the matrix is positive because of \eqref{Pauli2}.

However, fluctuation operators do indeed evolve under the mesoscopic dissipative dynamics that emerges from the microscopic one and their correlations are eventually embodied by the asymptotic mesoscopic state $\Omega_\infty$. Its covariance matrix $\Sigma_\infty$ is given by \eqref{asycovflu} with off-diagonal term 
$\displaystyle\Sigma_{12}(\infty)=-\frac{A_{12}}{2|b|}$: as it is not zero, it shows that quantum fluctuations become correlated by the mesoscopic dissipative dynamics and these correlations persist asymptotically in time.

Morover, the off-diagonal term originates from
\begin{eqnarray}
\nonumber
\Sigma_{12}(\infty)&:=&\frac{1}{2}
\lim_{t\to+\infty}\lim_{N\to\infty}\omega^{(N)}_t\left(\left\{F^{(N)}_t(s_1)\,,\,F^{(N)}_t(s_2)\right\}\right)\\
\nonumber
&=&
\lim_{t\to+\infty}\lim_{N\to\infty}\frac{1}{N}\sum_{i,j=1}^N\left(\omega^{(N)}_t\left(\frac{1}{2}\left\{s^{(i)}_1\,,\,s^{(j)}_2\right\}\right)\,-\,
\omega^{(N)}_t(s_1^{(i)})\,\omega^{(N)}_t(s_2^{(j)})\right)\\
\label{corrfluc}
&=&
\lim_{t\to+\infty}\lim_{N\to\infty}\frac{1}{N}\sum_{i\neq j=1}^N\left(\underbrace{\omega^{(N)}_t\left(\sigma_1^{(i)}\sigma_2^{(j)}\right)\,-\,
\omega^{(N)}_t(s_1^{(i)})\,\omega^{(N)}_t(s_2^{(j)})}_{C_{12}^{(ij)}(N,t)}\right)\ .
\end{eqnarray}
The last equality follows since $s_{1,2}$ at a same site anti-commute and from \eqref{invas}.
Since the state $\omega$ is a factor state, the mean-values of operators that are invariant under exchange of lattice sites are also invariant.
As the Kraus operators forming the Lindblad generator $\bL_N$ are also invariant under exchange of lattice indexes, it turns out that, 
$$
\omega^{(N)}_t\left(s^{(i)}_1\,s^{(j)}_2\right)=\omega^{(N)}_t\left(
s^{(k)}_1\,s^{(\ell)}_2\right)
\qquad \forall\ i\neq j\ ,\ k\neq\ell\ ,
$$
whence, for all $i\neq j$, \eqref{corrfluc} yields
$$
\lim_{t\to+\infty}\lim_{N\to\infty}N\left(C_{12}^{(ij)}(N,t)+\frac{A_{12}}{2N|b|}\right)=0\ .
$$
Then, we do not only know that the site-to-site correlations $C_{12}^{(ij)}(N,t)$ vanish with $N\to\infty$ at all times $t$, but also that, for $t\gg1$, they do it according to
\begin{equation}
C_{12}^{(ij)}(N,t)=-\frac{A_{12}}{2N|b|}+o(N^{-1},t)\ .
\end{equation}
This shows that the non-zero off-diagonal entries of the covariance matrix of the mesoscopic state $\Omega_\infty$ depends on correlations between local observables at the microscopic level that vanish in the large $N$ limit, but in a sufficiently slow manner that they can nevertheless contribute to mesoscopic correlations at the level of collective quantum fluctuations.

\section{Conclusions}

We considered a quantum spin chain consisting of two-level systems at each site and embedded within a common environment that gives rise to a microscopic dissipative, mean-field Lindblad type semigroup whose Kraus operators scale as the inverse square root of the number $N$ of sites. With respect to a translation invariant and clustering microscopic state, the large $N$ limit of such a dynamics when acting on local observables provides a one-parameter family of unitary automorphisms, this despite the microscopic dynamics being dissipative. Moreover, the dynamics is non-Markovian with a generator that depends on both final and initial time.

On the other hand, the dynamics of quantum fluctuations, namely quantum operators with zero microscopic mean that scale as the inverse square root of $N$, is a one-parameter family of completely positive maps that also break Markovianity in the same way as the
unitary dynamics on local observables.

Furthermore, the mesoscopic dissipative dynamics exhibits a stable scenario with convergence to a unique asymptotic Gaussian state with global correlations that have no microscopic correspondence and that survive the time asymptotic limit, and an unstable 
scenario where no asymptotic Gaussian state exists, due to the asymptotic divergence of any covariance matrix.

Though simple, the model studied here rather exhaustively shows the richness of possible situations that arise from a mean-field dissipative dynamics when their effects are studied on local operators or on operators that scale as fluctuations. Quantum fluctuations, their states and dynamics represent a mesoscopic level of description of many-body systems where
collective behaviours retain quantum footprints. As such, they appear very promising theoretical tools to model collective quantum behaviours at the interface of quantum and classical physics in a variety of physical contexts ranging from 
assemblies of nano-oscillators, quantum dots and ultra cold atoms, where the number of elementary constituents makes the presence of an environment and thus of external noise and dissipation hardly negligible.

\section{Appendix}

In this Appendix we provide a sketch of how to handle the large $N$ limit \eqref{derivation} and refer 
to \cite{BCFN} for the technical proof of the consistency of the following manipulations.
Let
\begin{equation}
\label{SigmaN}
\Sigma^{(N)}_{\mu\nu}(t):=\frac{1}{2}\Big\{F^{(N)}_t(s_\mu)\,,\,F_t^{(N)}(s_\nu)\Big\}\ .
\end{equation}
Then, the time-derivative of the mean-value of the above quantity with respect to the time-evolving state $\omega^{(N)}_t$ yields
\begin{eqnarray}
\nonumber
\frac{\rm {d}}{{\rm d}t}\omega^{(N)}_t\left(\Sigma^{(N)}_{\mu\nu}(t)\right)&=&\omega^{(N)}_t\left(\mathbb{L}_N\left[\Sigma^{(N)}_{\mu\nu}(t)\right]\right)\\
\nonumber
&+&\omega^{(N)}_t\left(\frac{1}{2}\left\{\dot{F}^{(N)}_t(s_\mu)\,,\,F^{(N)}_t(s_\nu)\right\}\,+\,\frac{1}{2}\left\{F^{(N)}_t(s_\mu)\,,\,\dot{F}^{(N)}_t(s_\nu)\right\}\right)\\ 
\label{derivation}
&=&\omega^{(N)}_t\left(\mathbb{L}_N\left[\Sigma^{(N)}_{\mu\nu}(t)\right]\right)\ .
\end{eqnarray}
Indeed, due to \eqref{timefluct2}, the contributions of the form 
$$
\dot{F}^{(N)}_t(s_\mu)F_t^{(N)}(s_\nu)=-\left(\frac{1}{\sqrt{N}}\sum_{k=1}^N\dot{\omega}^{(N)}_t(s^{(k)}_\mu)\right)\,F^{(N)}_t(s_\nu)\ ,
$$ 
have vanishing mean values with respect to $\omega^{(N)}_t$.

Let us first consider \eqref{Ag} with the collective operator $\Sigma^{(N)}_{\mu\nu}(t)$ in the place of $x^{(k)}$
\begin{equation}
\label{Ag2}
\bA_N\left[\Sigma^{(N)}_{\mu\nu}(t)\right]=\frac{1}{N}\sum_{k,\ell=1}^N\sum_{\mu',\nu'=1}^3\frac{A_{\mu\nu}}{2}\left[\left[s_{\mu'}^{(k)}\,,\,\Sigma^{(N)}_{\mu\nu}(t)\right]\,,\,s_{\nu'}^{(\ell)}\right]\ .
\end{equation} 
Because of the double commutator, the insertion of scalar quantities like the mean values 
$\omega^{(N)}_t(s^{(k)})$ does not alter the above expression that can thus be rewritten
\begin{equation}
\label{Ag3}
\bA_N\left[\Sigma^{(N)}_{\mu\nu}(t)\right]=\sum_{\mu',\nu'=1}^3\frac{A_{\mu'\nu'}}{2}\left[\left[F^{(N)}_t(s_{\mu'})\,,\,\Sigma^{(N)}_{\mu\nu}(t)\right]\,,\,F^{(N)}_t(s_{\nu'})\right]\ .
\end{equation}
Since commutators like $\left[F^{(N)}_t(s_{\mu'})\,,\,F^{(N)}_t(s_{\mu})\right]$ scale 
like mean-field observables, in the large $N$ limit they tend to the scalar quantities
$i\sigma^{\omega}_{\mu'\mu}(t)$ (see \eqref{Sympt}), so that, in the same limit,
\begin{equation}
\label{AN}
\left[F^{(N)}_t(s_{\mu'})\,,\,\Sigma^{(N)}_{\mu\nu}(t)\right]\simeq
i\,\sigma^\omega_{\mu'\mu}(t)\,F^{(N)}_t(s_\nu)\,+\,i\,\sigma^\omega_{\mu'\nu}(t)\,F^{(N)}_t(s_\mu)\ .
\end{equation}
Once inserted in \eqref{Ag3}, this behaviour yields
\begin{equation}
\label{Ag4}
\omega^{(N)}_t\left(\bA_N\left[\Sigma^{(N)}_{\mu\nu}(t)\right]\right)\simeq-\sum_{\mu',\nu'=1}^3\frac{A_{\mu'\nu'}}{2}\left(\sigma^\omega_{\mu'\mu}(t)\,\sigma^\omega_{\nu\nu'}(t)\,+\,\sigma^\omega_{\mu'\nu}(t)\,\sigma^\omega_{\mu\nu'}(t)\right)=\left(\sigma^\omega_t\,A\,(\sigma^\omega_t)^{tr}\right)_{\mu\nu}\ .
\end{equation}

Let us now consider the action 
\begin{equation}
\label{Bg2}
\bB_N\left[\Sigma^{(N)}_{\mu\nu}(t)\right]=\frac{i}{N}\sum_{k,\ell=1}^N\sum_{\mu',\nu'=1}^3\frac{B_{\mu'\nu'}}{2}\left\{\left[s_{\mu'}^{(k)}\,,\,\Sigma^{(N)}_{\mu\nu}(t)\right]\,,\,
s_{\nu'}^{(\ell)}\right\}\ .
\end{equation}
By inserting mean-values of the form $\omega^{(N)}_t(s_\mu^{(k)})$ in order to reconstruct fluctuation operators, it becomes
\begin{eqnarray}
\label{Bg3a}
\bB_N\left[\Sigma^{(N)}_{\mu\nu}(t)\right]&=&\frac{i}{N}\sum_{\mu',\nu'=1}^3\frac{B_{\mu'\nu'}}{2}\left\{\left[F_t^{(N)}(s_{\mu'})\,,\,\Sigma^{(N)}_{\mu\nu}(t)\right]\,,\,
F^{(N)}_t(s_{\nu'})\right\}\\
\label{Bg3b}
&+&
i\sum_{\mu',\nu'=1}^3B_{\mu'\nu'}\left(\frac{1}{\sqrt{N}}\sum_{k=1}^N\omega^{(N)}_t(s^{(k)}_{\nu'})\right)\,\left[F_t^{(N)}(s_{\mu'})\,,\,\Sigma^{(N)}_{\mu\nu}(t)\right]\ .
\end{eqnarray}

We now treat separately the contributions in \eqref{Bg3a} and \eqref{Bg3b} denoting the first one by
$\bB'_N\left[\Sigma^{(N)}_{\mu\nu}(t)\right]$ and by $\bB''_N\left[\Sigma^{(N)}_{\mu\nu}(t)\right]$ the second one.
Using \eqref{AN}, in the large $N$ limit one gets
\begin{eqnarray}
\nonumber
\omega^{(N)}_t\left(\bB'_N\left[\Sigma^{(N)}_{\mu\nu}(t)\right]\right)&\simeq&
-\sum_{\mu',\nu'=1}^3B_{\mu'\nu'}\left(\sigma^\omega_{\mu'\mu}(t)\,\Sigma_{\nu\nu'}(t)\,+\,\sigma^\omega_{\mu'\nu}(t)\,\Sigma_{\mu\nu'}(t)\right)\\
\label{Bg4a}
&=&\Big(\sigma^\omega_t\,B\,\Sigma_t\,+\,\Sigma_t\,B\,\sigma^\omega_t\Big)_{\mu\nu}\ ,
\end{eqnarray}
where $\sigma^\omega_{\mu\nu}(t)=-\sigma^\omega_{\nu\mu}(t)$,
$B_{\mu\nu}=-B_{\nu\mu}$ and $\Sigma_{\mu\nu}(t)=\Sigma_{\nu\mu}(t)$ have been used.

In order to control the large $N$ limit of the mean-value of $\bB''_N\left[\Sigma^{(N)}_{\mu\nu}(t)\right]$, notice that
\begin{eqnarray*}
\left[F_t^{(N)}(s_{\mu'})\,,\,F^{(N)}_t(s_\mu)\right]&=&\frac{1}{N}\sum_{k,\ell=1}^N
\Big[s^{(k)}_{\mu'}\,,\,s^{(\ell)}_\mu\Big]=i\epsilon_{\mu'\mu\gamma}\,\frac{1}{N}
\sum_{k=1}^Ns^{(k)}_\gamma\\ 
&=&i\epsilon_{\mu'\mu\gamma}\,\frac{1}{\sqrt{N}}\,F^{(N)}_t(s_\gamma)\,+\,
i\epsilon_{\mu'\mu\gamma}\,\frac{1}{N}
\sum_{k=1}^N\omega^{(N)}_t(s^{(k)}_\gamma)\ .
\end{eqnarray*}
Then,
\begin{eqnarray*}
&&\hskip-.5cm
\left[F_t^{(N)}(s_{\mu'})\,,\,\Sigma^{(N)}_{\mu\nu}(t)\right]=
\frac{i}{\sqrt{N}}\epsilon_{\mu'\mu\gamma}\Sigma^{(N)}_{\gamma\nu}(t)\,+\,\frac{i}{\sqrt{N}}\epsilon_{\mu'\nu\gamma}\,\Sigma^{(N)}_{\gamma\mu}(t)\\
&&\hskip 1cm
+\,
i\,\epsilon_{\mu'\mu\gamma}\,\left(\frac{1}{N}\sum_{k=1}^N\omega^{(N)}_t(s^{(k)}_\gamma)\right)\,F^{(N)}_t(s_\nu)\,+\,i\epsilon_{\mu'\nu\gamma}\,\left(\frac{1}{N}\sum_{k=1}^N\omega^{(N)}_t(s^{(k)}_\gamma)\right)\,F^{(N)}_t(s_\mu)\ .
\end{eqnarray*}
Since fluctuation operators have zero mean-values with respect to the state 
$\omega^{(N)}_t$, one finally gets
$$
\omega^{(N)}_t\left(\bB''_N\left[\Sigma^{(N)}_{\mu\nu}(t)\right]\right)=
-\sum_{\mu',\nu'=1}^3B_{\mu'\nu'}\left(\frac{1}{N}\sum_{k=1}^N\omega^{(N)}_t(s_{\nu'})\right)\,\left(\epsilon_{\mu'\mu\gamma}\Sigma^{(N)}_{\gamma\nu}(t)\,+\,\epsilon_{\mu'\nu\gamma}\,\Sigma^{(N)}_{\gamma\mu}(t)\right)\ ,
$$
whence, in the large $N$ limit, where $\sum_{k=1}^N\omega^{(N)}_t(s^{(k)}_{\nu'})/N\to\omega_{\nu'}(t)$,
\begin{eqnarray}
\nonumber
\omega^{(N)}_t\left(\bB''_N\left[\Sigma^{(N)}_{\mu\nu}(t)\right]\right)&\simeq&
-\sum_{\mu',\nu'=1}^3B_{\mu'\nu'}\,\omega_{\nu'}(t)\,\Big(\epsilon_{\mu'\mu\gamma}\Sigma_{\gamma\nu}(t)\,+\,\epsilon_{\mu'\nu\gamma}\,\Sigma_{\gamma\mu}(t)\Big)\\
\label{Bg5a}
&=&\left(C_t\,\Sigma_t\,+\,\Sigma_t\,C^{tr}\right)_{\mu\nu}\ ,
\end{eqnarray}
where $C_t$ is the anti-symmetric matrix with entries
\begin{eqnarray}
\label{Bg5b}
C_{\mu\nu}(t)=\sum_{\mu',\nu'=1}^3\epsilon_{\mu\mu'\nu}\,B_{\mu'\nu'}\,\omega_{\nu'}(t)
\ .
\end{eqnarray}
Putting together \eqref{Ag4}, \eqref{Bg4a} and \eqref{Bg5a} one finally gets the result in \eqref{Smateq}.

\bibliographystyle{plain}

\end{document}